# Reversible Self-Assembly of a P-Ir(100)-P Trilayer on Ir(111)

M. Bassotti,[a,*] J. Sivianes,[a] J. Dai,[b] M. Tallarida,[b] P. Angulo-Portugal,[a] A.A. Makarova,[c] J. Ibañez-Azpiroz,[a,d,e] D.Yu. Usachov,[f,g] A. Verdini,[h] and F. Schiller [a,*].

*a) Centro de Física de Materiales CSIC/UPV-EHU – Materials Physics Center, 20018 San Sebastián, Spain*

*b) ALBA Synchrotron Light Source, Cerdanyola del Vallès, 08290 Barcelona, Spain*

*c) Helmholtz-Zentrum Berlin für Materialien und Energie, BESSY II, Albert-Einstein-Str 15, 12489 Berlin, Germany*

*d) IKERBASQUE, Basque Foundation for Science, 48009 Bilbao, Spain*

*e) Donostia International Physics Center (DIPC), 20018 Donostia-San Sebastián, Spain*

*f) St. Petersburg State University, 7/9 Universitetskaya nab., 199034 St. Petersburg, Russia*

*g) Moscow Institute of Physics and Technology, Institute Lane 9, 141701 Dolgoprudny, Russia*

*h) Consiglio Nazionale delle Ricerche – Istituto Officina dei Materiali (IOM), 06123 Perugia, Italy*

** mattia.bassotti@ehu.eus, frederikmichael.schiller@ehu.eus*

## Abstract:

Adsorbate-induced surface reconstructions offer a route to engineer low-dimensional materials with atomic arrangements and electronic properties fundamentally different from those of their parent substrates. Here, we demonstrate how phosphorus adsorption on Ir(111) drives the formation of an incommensurate P–Ir–P trilayer, in which a square Ir(100)-like atomic lattice emerges from the underlying hexagonal crystal. This transformation illustrates how strong adsorbate–substrate interactions can stabilize unconventional atomic coordination within an ultrathin metallic structure. Remarkably, the process is reversible, allowing the system to switch between distinct structural phases through phosphorus deposition and thermal annealing. The resulting trilayer exhibits a two-dimensional, semimetal-like electronic structure dominated by Ir 5d states. Our findings establish adsorbate-driven symmetry transformation as a strategy for engineering atomically thin transition-metal compounds with structural and electronic properties inaccessible through conventional surface growth.

## Introduction:

Adsorbate–substrate interactions and low-dimensional growth can drive mass transport, reconstruction, and faceting, stabilizing surface structures whose geometry and symmetry differ from the bulk-truncated crystal [1–4], opening the way for nanoscale engineering of surfaces [5].

A particularly striking case occurs when adsorption locally transforms the close-packed hexagonal termination of an fcc(111) metal into an expanded, distorted square lattice with (100)-like coordination. Carbon- and sulfur-containing species on Ni(111), for example, produce multiple rotational domains of near-square structures related to Ni(100)-(2 × 2), with the reconstruction linked to the relaxation of adsorbate-induced compressive stress [6,7]. Similarly, oxygen adsorption and incorporation on Cu(111) generate distorted Cu(100)-like domains, while methylthiolate induces a thermally activated pseudo-Cu(100) reconstruction in which the S head groups acquire predominantly near-fourfold coordination [8,9]. Microscopically, these transformations arise from strong hybridization between adsorbate *p* states and metal *d* bands, which redistributes charge within the surface layer, weakens lateral metal–metal bonding, and generates compressive stress, thereby favoring higher coordination in a square-like geometry [7].

Phosphorus likewise strongly hybridizes through its 3*p* states with transition-metal *d* states and readily forms surface phosphides and metal–phosphorus hybrid phases on reactive substrates [10–14], making it a natural candidate to stabilize an analogous square, (100)-like metal arrangement. Its growth can also reorganize the substrate over larger length scales: on curved Cu, for instance, phosphorus induces Cu(513) facets that selectively stabilize a square phosphorus phase [5]. Ir–P chemistry provides a particularly appealing platform for such coordination engineering. Bulk REIr2P2 compounds contain tetragonal [Ir2P2] networks with transition-metal squares and strong Ir–P bonding [15], demonstrating that phosphorus can stabilize square structural motifs involving Ir. To our knowledge, however, such a local conversion of the triangular Ir(111) geometry into a square arrangement has not been realized in an atomically thin Ir–P phase.

Here, we demonstrate the growth of an incommensurate square-like P–Ir–P trilayer on Ir(111), composed of a central Ir sheet sandwiched between two square phosphorus layers. The Ir plane has the geometry and nearest-neighbor spacing of an Ir(100) layer despite being supported on the hexagonal Ir(111) surface. We further identify a phosphorus–iridium surface alloy as a precursor to the trilayer and show that the transformation between the two phases is reversible: additional phosphorus drives trilayer formation, whereas thermal annealing restores the alloy phase. These results establish phosphorus incorporation as a route to reversibly engineer the symmetry, coordination, and electronic structure of an atomically thin 5d-metal compound.

## Results and Discussion:

Initial deposition of phosphorus on Ir(111) produces the low-energy electron diffraction, or LEED, pattern shown on the left of Fig. 1(a). The overlayer forms a rectangular $2\times\sqrt{3}$-R30° superstructure with respect to Ir(111), described in matrix notation as $\begin{pmatrix}2 & 1\\ 0 & 2\end{pmatrix}$, occurring in three rotational domains. Taking the Ir(111) nearest-neighbor distance as 2.71 Å, the corresponding rectangular unit cell has dimensions of 5.42 Å and 4.70 Å. Although the aim of the present work is not on the study of this structure, we suggest the structural model that is shown in Fig. S1, with a single P atom per unit cell, in a mixture of either fcc or hcp position with respect to the Ir(111) substrate. The validity of the model is supported by comparison between the measured and calculated Photoemission Electron Diffraction (PED) shown in Fig. S2.

Upon further phosphorus deposition, this phase is replaced by the LEED pattern shown in the central panel of Fig. 1(a). The overlayer now displays a square unit cell, again occurring in three rotational domains, while the Ir(111) diffraction spots are no longer visible. The square lattice parameter is approximately 4 Å, as determined using the Ir(111) reciprocal lattice visible on the preceding phase as a reference. Additional satellite spots are observed around the specular (0,0) beam, indicating the formation of a long-period moiré modulation. Together, these observations are consistent with an overlayer that is incommensurate with the underlying Ir(111) surface.

The measured square lattice parameter is close to the bulk Ir lattice constant, $a_{Ir}$ = 3.84 Å. Because a purely phosphorus square lattice with this periodicity would be difficult to reconcile with the strong P–Ir interaction, we considered a structure in which Ir atoms participate directly in the overlayer. A compatible structural model is shown in Fig. 1(b), with top and side views displayed on the left and right, respectively. The relaxed structure consists of an Ir layer arranged in an Ir(100)-like square geometry and sandwiched between two P layers. Each P layer forms a $\sqrt{2}\times\sqrt{2}$-R45° structure with respect to the Ir(100)-like layer. The resulting surface unit cell has a side length matching with the bulk Ir lattice constant and is therefore consistent with the periodicity extracted from LEED.

Further support for this assignment is provided by phosphorus deposition directly on a Ir(100) substrate. The corresponding LEED pattern, shown on the right-hand side of Fig. 1(a), displays a $\sqrt{2} \times \sqrt{2}$-R45° structure, in agreement with the local atomic arrangement proposed for the P–Ir–P trilayer. Remarkably, rather than forming a merely pseudosquare overlayer on the hexagonal Ir(111) surface as seen in analogous systems [6,9], phosphorus adsorption drives the emergence of an actual Ir(100)-like atomic layer with the bulk Ir lattice parameter.

A further structural feature of the trilayer is illustrated in Fig. 1(c). For simplicity, the top view shows only the P atoms of the atomic model, together with their heights relative to the Ir(111) substrate. The P atoms are not located at a uniform height but exhibit a pronounced one-dimensional modulation along a direction rotated by 45° with respect to the square unit-cell vectors, indicated in red. The calculated modulation period, marked by the two parallel red lines, is of 1.9 nm. This corrugation is directly reflected in the room-temperature scanning tunneling microscopy, or RT-STM, image and height profile shown in Fig. 1(d) (see Fig. S3 for a large-scale STM image). The green dashed line follows an undulating pattern oriented at 45° with respect to the square trilayer unit cell, outlined in black. The periodicity extracted from STM is approximately 1.7 nm. The small difference from the calculated value can be attributed to thermal drift in the RT-STM measurements, which are also evident from the slight distortion of the atomic lattice. The measured corrugation amplitude, approximately 70 pm, is also in good agreement with the 55 pm modulation predicted by the relaxed structural model, further supporting the proposed geometry. Finally, the enlarged LEED image around the specular beam, shown in the inset of Fig. 1(d), reveals a sixfold arrangement of moiré satellites corresponding to a real-space periodicity of approximately 1.9 nm. This pattern can be understood as the superposition of the one-dimensional height modulations associated with the three rotational domains of the P–Ir–P trilayer.

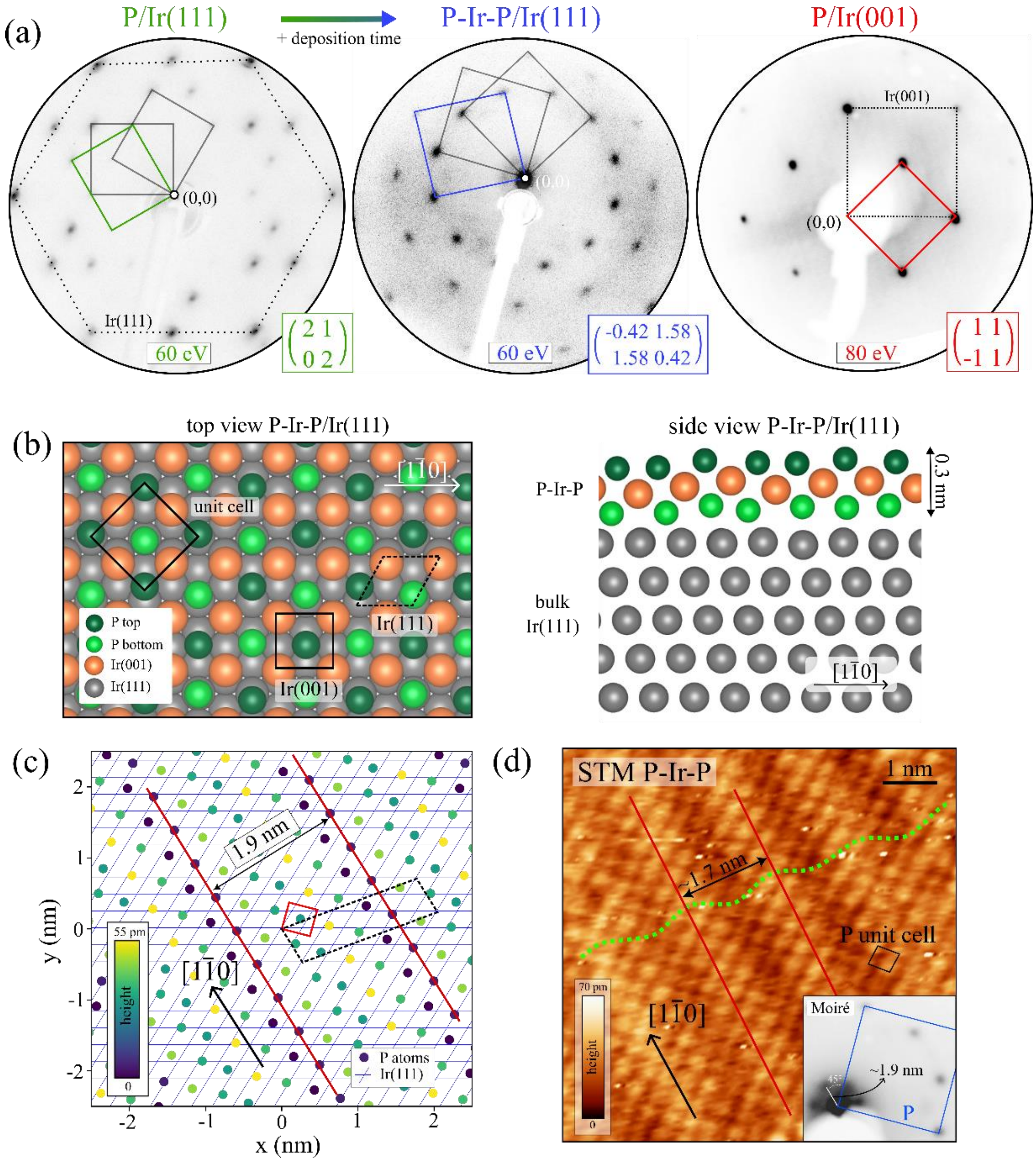

**Figure 1. Structural characterization of the P/Ir phases and P–Ir–P trilayer model.** (a) LEED patterns of the precursor P/Ir(111), P–Ir–P/Ir(111), and P/Ir(100) phases, acquired at electron energies of 60, 60, and 80 eV, respectively. The corresponding surface unit cells are highlighted in color, while the substrate unit cell, where visible, is indicated by the black dashed outline. The reconstruction matrices are also reported. (b) Relaxed structural model of the P–Ir–P trilayer shown in top (left) and side (right) views. The top view highlights the unit cells of the P overlayer, the embedded Ir(100)-like layer, and the Ir(111) substrate. The side view shows the trilayer thickness. (c) Height profile of the P atoms obtained from the relaxed structure. The red lines indicate the calculated corrugation period of 1.9 nm. The P unit cell is outlined in red, while the DFT supercell is indicated by the black dashed outline. (d) Room-temperature STM image of a single rotational domain of the P–Ir–P trilayer. The red lines mark the measured corrugation period of approximately 1.7 nm, oriented 45° with respect to the square P unit-cell vectors (imaging parameters: 7x7 nm$^2$, I = 200 pA, bias = 1 V). Inset: Enlarged LEED image around the specular (0,0) beam showing the moiré satellites. The P unit cell and the moiré periodicity are reported. LEED energy = 60 eV.

The X-Ray Photoelectron Spectroscopy (XPS) characterization of the precursor and trilayer phases is presented in Fig. 2(a). The P 2p spectrum of the low-coverage precursor phase, shown in the upper panel, is well described by a dominant spin–orbit doublet, depicted in green, with the P $2p_{3/2}$ component centered at 128.78 eV. Following further P deposition, the spectral line shape changes substantially, yielding the spectrum shown in the lower panel of Fig. 2(a). The trilayer spectrum requires two main spin–orbit doublets, shown in dark and light blue, with P $2p_{3/2}$ binding energies of 129.25 and 129.93 eV, respectively. The presence of two chemically distinct P components is consistent with the two inequivalent P layers of the proposed P–Ir–P trilayer.

The shift of approximately 0.5 eV between the precursor component and the lower-binding-energy component of the trilayer indicates a pronounced modification of the local electronic environment of P. We attribute this change to the different P–Ir coordination in the two structures and, in particular, to the formation of stronger P–Ir bonds within the Ir(100)-like trilayer. This enhanced interaction may contribute to stabilizing the square P–Ir–P structure on Ir(111). Weak additional components, represented by the black areas, are assigned to minority P species associated with structural defects, domain boundaries, or residual disorder.

The integrated P 2p intensities provide an additional consistency check for the structural model. The calculated ratio of P areal densities for the two phases is 3.5, in reasonable agreement with the corresponding experimental P 2p intensity ratio of 3.0. A strictly proportional relationship between XPS intensity and P areal density is not expected in this case. Indeed, in addition to the different number of P atoms per unit area, the measured intensity is affected by photoelectron-diffraction and matrix-element effects, which depend on the local atomic geometry and emission direction. Furthermore, photoelectrons emitted from the lower P layer of the trilayer are partially attenuated while passing through the overlying Ir and P atoms. Considering these effects, the agreement between the calculated density ratio and the measured intensity ratio provides further support for the proposed trilayer structure.

The phase transformation induced by additional P deposition raises the question of whether the process is reversible, namely whether the precursor phase can be recovered by removing P from the trilayer through thermal annealing. Fig. 2(b) shows the temperature-dependent evolution of the P 2p core level during annealing of the P–Ir–P trilayer from 700 to approximately 1040 K. The lower dashed white line corresponds to the spectrum recorded at T = 750 K and reported in Fig. S4, confirming the presence of the trilayer at the beginning of the temperature ramp.

The trilayer remains stable up to approximately 950 K, above which P desorption and a progressive transformation of the P 2p line shape are observed. At higher temperatures, one of the two trilayer-related doublets gradually disappears, while the remaining component shifts toward lower binding energy. The conversion is essentially complete above approximately 1025 K. The final spectrum, measured at 1030 K and shown in Fig. S4, closely reproduces the P 2p line shape obtained after the initial P-deposition step, demonstrating the recovery of the precursor phase.

Combined with the two-step deposition experiment, the temperature-dependent measurements establish the reversibility of the P–Ir–P trilayer formation. At lower P coverage, the rectangular $2\times\sqrt{3}$ precursor overlayer is stabilized on Ir(111). Increasing the amount of available P drives a structural rearrangement in which Ir atoms are incorporated into a square P–Ir–P trilayer. Conversely, thermal removal of P converts the trilayer back into the precursor structure. These measurements also reveal the remarkable thermal stability of both phases: the trilayer remains intact up to approximately 950 K, whereas the precursor persists up to the highest temperature investigated, 1040 K.

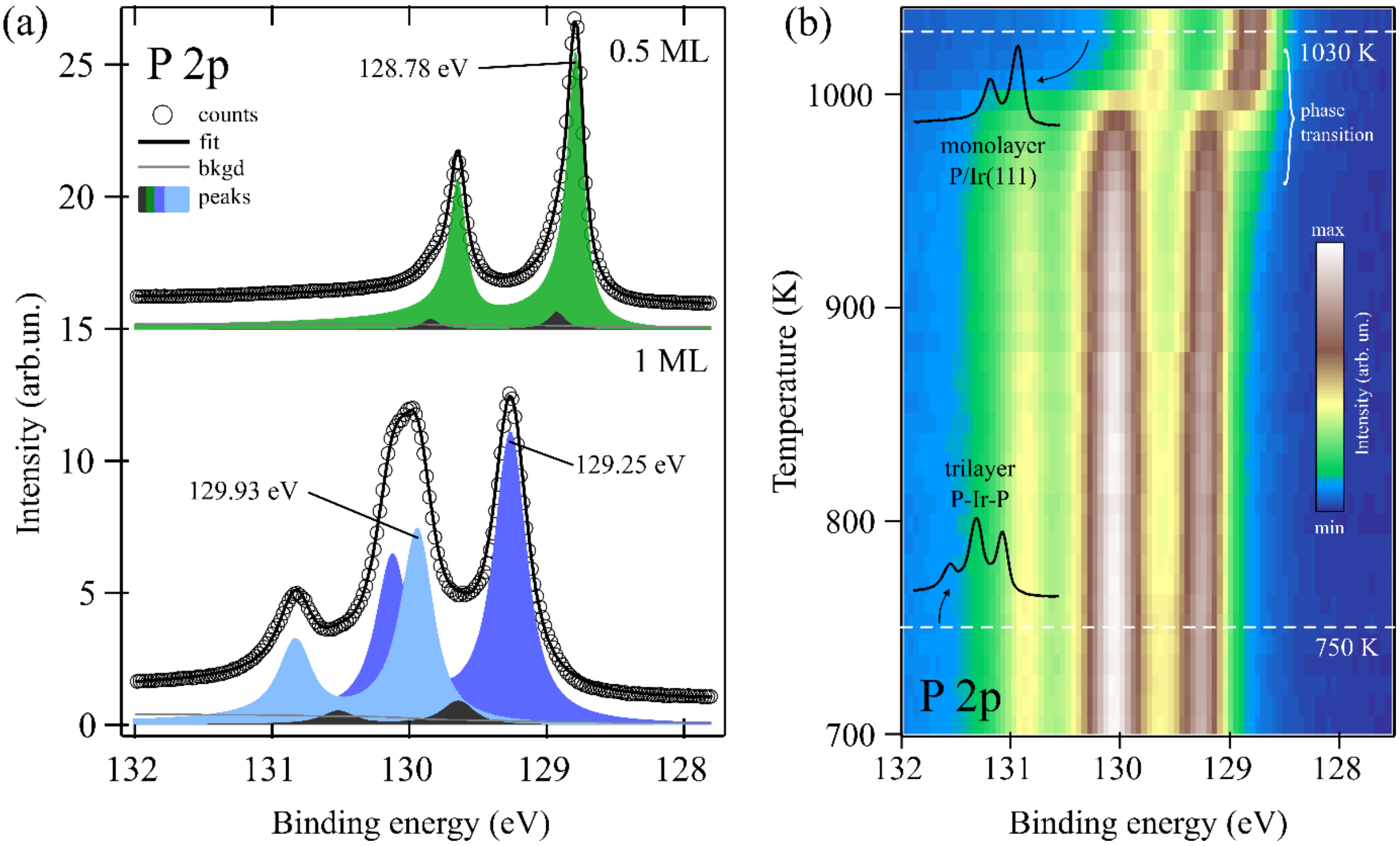

**Figure 2. XPS characterization and thermal evolution of the P–Ir–P trilayer.** (a) P 2p XPS spectra of the precursor P/Ir(111) phase (top) and the P–Ir–P trilayer (bottom). The binding energies of the main peaks are shown. Photon energy: 250 eV. (b) Temperature-dependent evolution of the P 2p core level during annealing of the P–Ir–P trilayer. Representative spectra acquired at 750 K and 1030 K, corresponding to the trilayer and precursor phases, respectively, are also shown. The approximate temperature range over which the phase transformation occurs is indicated. Photon energy: 230 eV.

We now discuss the electronic properties of the trilayer phase, presented in Fig. 3. In particular, Fig. 3(a) shows the ARPES intensity measured along the Γ–M direction of the suibstrate using linear-vertical (LV) and linear-horizontal (LH) light polarization, displayed in the left and right panels, respectively. The corresponding experimental geometry is illustrated in the sketch on the right. The *xz* plane defines the measurement plane. In the LV configuration, the electric field lies parallel to the sample surface and perpendicular to the measurement plane, and is therefore directed along *y*. By contrast, in the LH configuration, the electric field lies within the measurement plane and has finite *x* and *z* components. Assuming that the final-state photoelectron wave function is even with respect to reflection across the *xz* plane, the two polarizations selectively probe initial states with different mirror parities: LH primarily couples to states that are even with respect to the *xz* plane, whereas LV probes odd states. The interpretation of this polarization dependence is, however, complicated by the presence of three rotational domains of the trilayer, which mix contributions from orbitals with different in-plane orientations.

The ARPES spectra reveal three main dispersive features between the Fermi level and approximately 5 eV binding energy. All three display hole-like dispersions centered at Γ. The uppermost bands approach, and within the experimental resolution appear to reach, the Fermi level, indicating a semimetal-like electronic structure with a cone-like dispersion reminiscent of graphene. These features are considerably more intense in LH than in LV polarization, suggesting that they predominantly originate from states with even symmetry with respect to the *xz* plane. Conversely, the band centered at approximately 2 eV binding energy at Γ is relatively enhanced in LV polarization, indicating a larger contribution from odd-symmetry states. Because of the coexistence of the three rotational domains, however, these polarization dependences cannot be converted into an unambiguous assignment of individual in-plane orbitals.

The momentum-space evolution of the uppermost bands is further illustrated by the constant-energy maps in Fig. 3(b). Down to a binding energy of approximately 1.6 eV, the photoemission intensity forms nearly concentric contours around Γ, consistently with the hole-like dispersions observed in the energy–momentum cuts. The contours remain approximately isotropic over this energy range. This apparent isotropy may be partly enhanced by the superposition of the three rotational domains, which averages the in-plane anisotropy of the individual domains.

The two-dimensional character of these states is supported by the photon-energy-dependent measurements reported in Fig. S5. The principal dispersive features associated with the trilayer persist over the investigated photon-energy range without appreciable changes in their in-plane momentum dispersion. The absence of a pronounced photon-energy dependence indicates negligible $k_z$ dispersion and supports their assignment to states predominantly localized within the trilayer rather than to bulk-like states of the Ir substrate. The calculated band structure is presented in Fig. 3(c). To identify the orbital origin of the experimentally observed features, the calculated bands were projected onto the Ir 5d orbitals.

For a single trilayer domain aligned with the experimental coordinate system, the calculated orbital contributions were grouped according to their parity with respect to the *xz* plane. The LH-like projection was obtained by summing the even $d_{z2}$, $d_{x2-y2}$ and $d_{xz}$ contributions, whereas the LV-like projection was obtained by summing the odd $d_{xy}$ and $d_{yz}$ contributions. The red dashed lines in Fig. 3(a) and Fig. 3(c) are guides to the eye highlighting the dispersive features that can be identified in both the experimental spectra and the calculations, and facilitate their direct comparison.

The orbital-resolved calculations reproduce the main polarization-dependent trends observed experimentally. In the LH-like projection, the upper hole-like bands responsible for the concentric contours around Γ carry substantial spectral weight, whereas their intensity is strongly reduced in the LV-like projection. Conversely, the band centered near 2 eV binding energy at Γ has a comparatively larger contribution from odd orbitals, in agreement with its enhanced visibility in the experimental LV spectrum.

A rigid energy offset is observed between the experimental and calculated band structures. Nevertheless, the relative energy separations and dispersions of the principal features are largely preserved, allowing the corresponding bands to be identified despite the shift. The calculations also predict several additional bands that are weak or absent in the experimental spectra. Some of these states originate from the underlying Ir(111) substrate, while others may arise from band folding associated with the large supercell used to model the trilayer/substrate system. Experimentally, the strong surface sensitivity of ARPES substantially attenuates photoemission from the buried Ir substrate. In addition, photoemission matrix-element effects can suppress the intensity of folded and substrate-derived states, accounting for the lower number of bands observed experimentally.

Finally, the calculations indicate a region with no trilayer-derived electronic states immediately above the uppermost Γ-centered band. This suggests that an electronic gap may open just above the present Fermi level. If the chemical potential could be shifted so that the uppermost band moves to higher binding energy, for example through controlled doping, the Fermi level might be brought into this gap, potentially driving the trilayer from its present semimetal-like state toward a more semiconductor-like electronic configuration. Although this scenario remains speculative, it suggests a possible route for future control of the electronic properties of the trilayer.

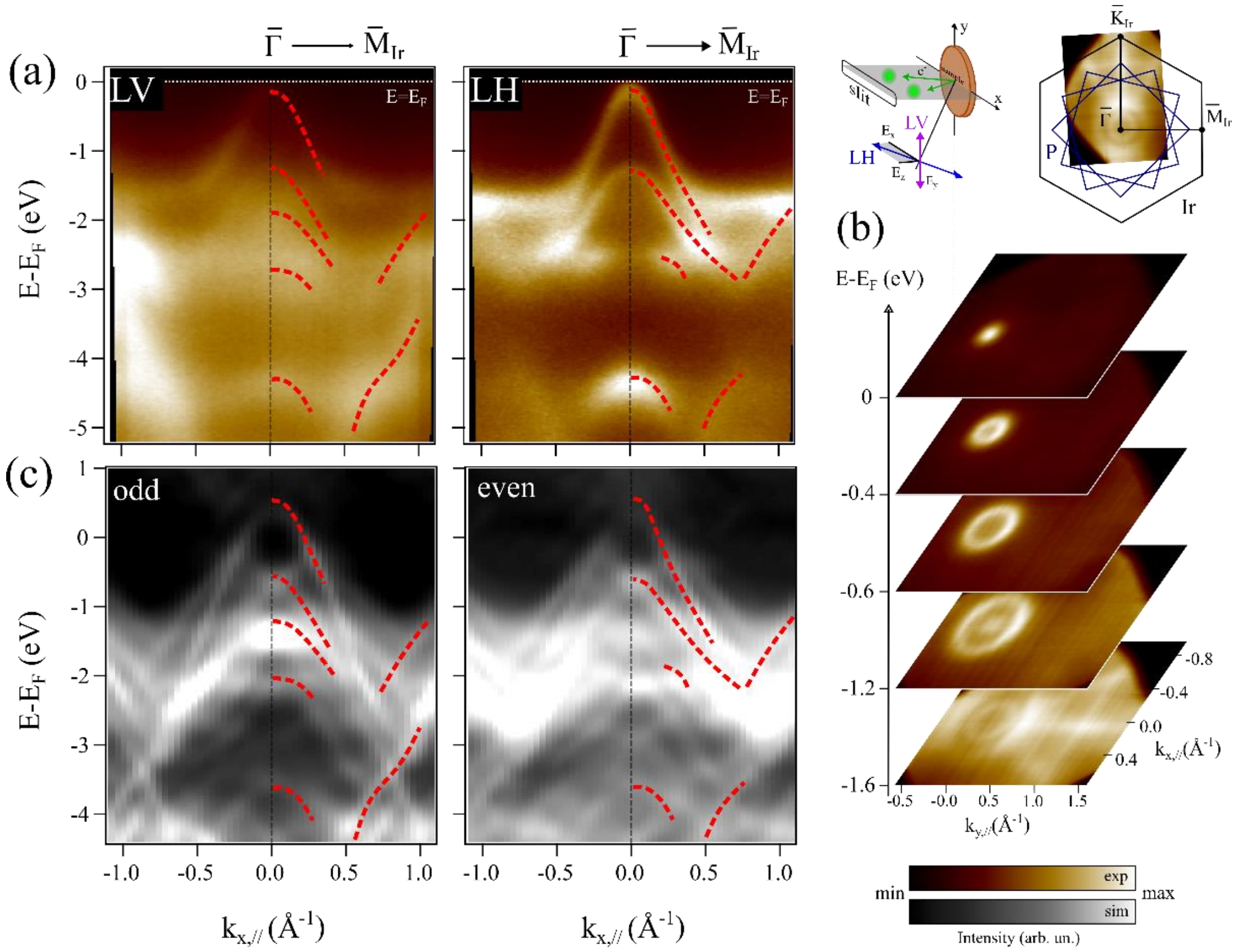

**Figure 3. Trilayer Angle-Resolved Photoemission Spectroscopy measurements.** (a) Measured band structure of the P-Ir-P trilayer on Ir(111) along the Γ–$M_{Ir}$ direction for the LV and the LH polarizations, which geometries are sketched on the right-hand side. The main features are highlighted by the red dashed lines. Photon energy = 72 eV. (b) Constant energy maps as a function of the binding energy using LH polarization. Photon Energy = 72 eV. On top, the Brillouin zones of the three P rotational domains and Ir(111) substrate are sketched. (c) Calculated band structure along Γ–$M_{Ir}$. The main feature of the measured band structure in (a) are also reported.

## Conclusions:

In summary, we have demonstrated the reversible formation of an atomically thin P–Ir–P trilayer on Ir(111), in which phosphorus stabilizes a square Ir(100)-like atomic arrangement on a hexagonal substrate. The combined structural characterization and theoretical calculations reveal an incommensurate trilayer exhibiting a long-period moiré modulation, while temperature-dependent XPS establishes its reversible transformation into a lower-coverage precursor phase through phosphorus desorption. The trilayer remains stable up to approximately 950 K, highlighting the robustness of the reconstructed structure. ARPES measurements, supported by orbital-resolved calculations, further reveal a two-dimensional, semimetal-like electronic structure with hole-like bands approaching the Fermi level. Beyond the formation of a distinct two-dimensional Ir–P compound, these findings demonstrate how adsorbate-induced reconstruction can transform the local coordination and symmetry of a metal surface, providing a route toward the controlled stabilization of low-dimensional structures with electronic properties distinct from those of the parent substrate.

## Methods:

*Sample Preparation.*

All sample preparation were carried out under ultra-high vacuum (UHV) conditions with a base pressure below $5\times10^{-10}$ mbar. Ir(111) and Ir(100) single crystals (Mateck, 99.99% purity) were cleaned by repeated cycles of $Ar^+$ sputtering (1 kV) followed by annealing at 1200 K until sharp LEED patterns characteristic of the clean surfaces were obtained. Phosphorus was deposited by thermal evaporation of red phosphorus (Merck, > 99.99 % purity) from a resistively heated crucible maintained at 320 °C, as monitored by a thermocouple attached to the crucible. During evaporation, the phosphorus flux was continuously monitored with a quadrupole mass spectrometer by tracking the $P_1$ (31 amu) and $P_2$ (62 amu) signals, which were maintained at approximately $5\times10^{-10}$ mbar. Throughout the deposition, the substrate temperature was kept at 450 °C. For the Ir(111) substrate, deposition for 10 min produced the precursor P/Ir(111) phase, while an additional 10 min of deposition under identical conditions resulted in the formation of the P–Ir–P trilayer. For the Ir(100) substrate, the same evaporation conditions were employed, using a deposition time of 10 min. After deposition, the sample temperature was maintained at 450 °C for an additional 10 min with the phosphorus source switched off.

*Experimental Techniques.*

All experiments were conducted in UHV conditions, with pressure better than $5 \times 10^{-10}$ mbar.

Scanning Tunneling Microscopy (STM) measurements were performed at room temperature using a commercial variable-temperature STM (Scienta Omicron, Centro de Física de Materiales, San Sebastián, Spain) equipped with chemically etched tungsten tips.

The X-ray photoelectron spectroscopy (XPS) measurements shown in Fig. 2(a) and the angle-resolved photoemission spectroscopy (ARPES) data presented in Fig. 3(a) were acquired at the LOREA beamline of the ALBA Synchrotron (Barcelona, Spain). A MB Scientific A-1 hemispherical electron analyzer with an angular resolution of 0.2° was employed. The P 2p XPS spectra were recorded using a photon energy of 250 eV. The spectra were fitted using Doniach–Šunjić line shapes for the P 2p spin–orbit doublets together with a Shirley background. ARPES measurements were performed at a photon energy of 72 eV and a sample temperature of 120 K.

The temperature-dependent XPS measurements shown in Fig. 2(b) were carried out at the GELEM end station of the BESSY II synchrotron (Berlin, Germany). The P 2p spectra were acquired using a photon energy of 230 eV. During the measurements, the sample temperature was increased linearly at a rate of 5 K $min^{-1}$ while P 2p spectra were continuously recorded. All binding energies were calibrated with respect to the measured Fermi level.

*Computational Details.*

Fully relativistic density functional theory (DFT) calculations were performed using the Vienna Ab initio Simulation Package (VASP) [16,17], with projector augmented-wave (PAW) pseudopotentials [18] and the Perdew–Burke–Ernzerhof (PBE) exchange–correlation functional [19].

The system consisted of a P–Ir–P trilayer supported on an Ir(111) substrate. The trilayer has a square unit cell with lattice vectors $a_1 = (a, 0)$ and $a_2 = (0, a)$, where a = 3.85 Å is the Ir lattice constant. The 15° rotation of the trilayer relative to the substrate renders the structure incommensurate. A commensurate approximation was constructed by applying a uniform isotropic compression of 0.5% to the trilayer, yielding the supercell vectors $A_1 = (2.72, -4.72)$ Å and $A_2 = (17.71, 7.08)$ Å.

A plane-wave energy cutoff of 600 eV and a Γ-centered $9 \times 3 \times 1$ k-point mesh were used. All atomic positions were relaxed until the residual forces were below $10^{-2}$ eV/Å, with an electronic convergence tolerance of $10^{-7}$ eV. Gaussian smearing with a width of 0.05 eV was applied.

## Acknowledgments:

# SUPPLEMENTARY INFORMATION

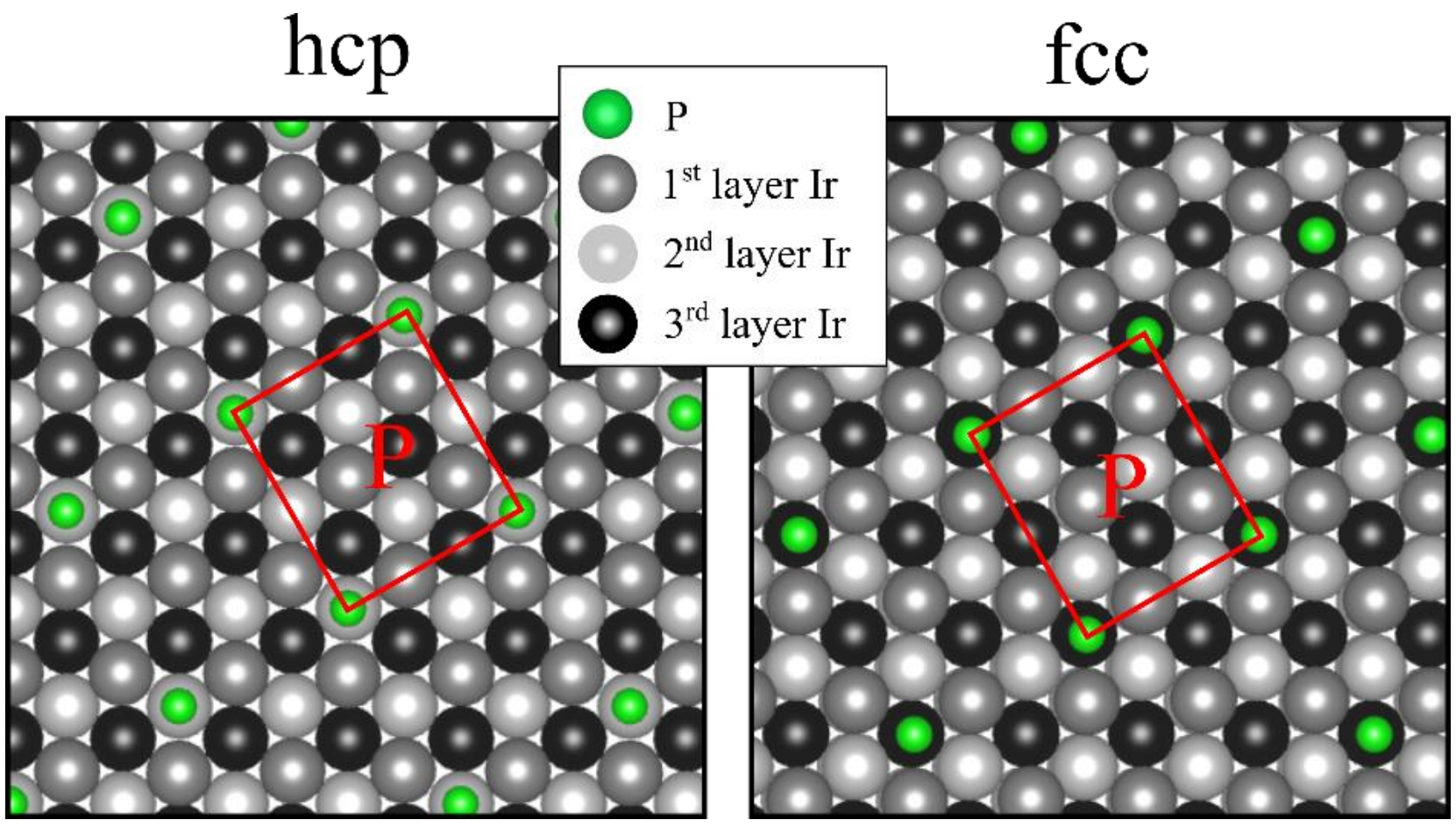

**Figure S1. Structural models of the initial stage of P deposition on Ir(111).** Top-view representations of the $(2\times\sqrt{3})$-R30° P superstructure on Ir(111), showing two possible adsorption configurations in which the P atoms occupy either hcp or fcc hollow sites relative to the underlying Ir lattice.

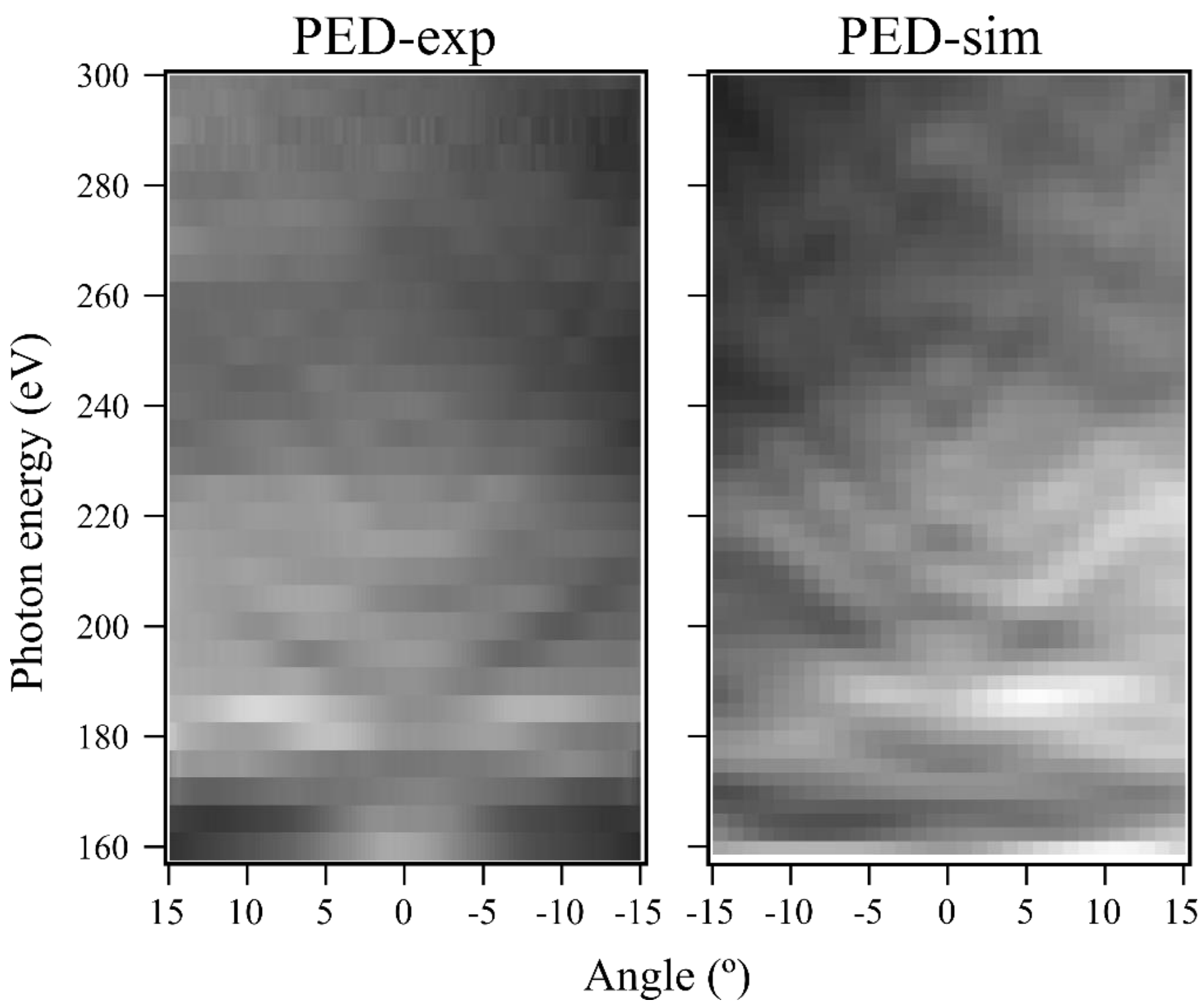

**Figure S2. Experimental and simulated PED patterns.** Left: Experimental PED pattern obtained by measuring the P 2p core level as a function of photoelectron emission angle and photon energy, varied from 160 to 300 eV in 5 eV steps. Right: Simulated PED pattern based on the structural models shown in Fig. S1, assuming a mixture of 70% fcc and 30% hcp adsorption configurations.

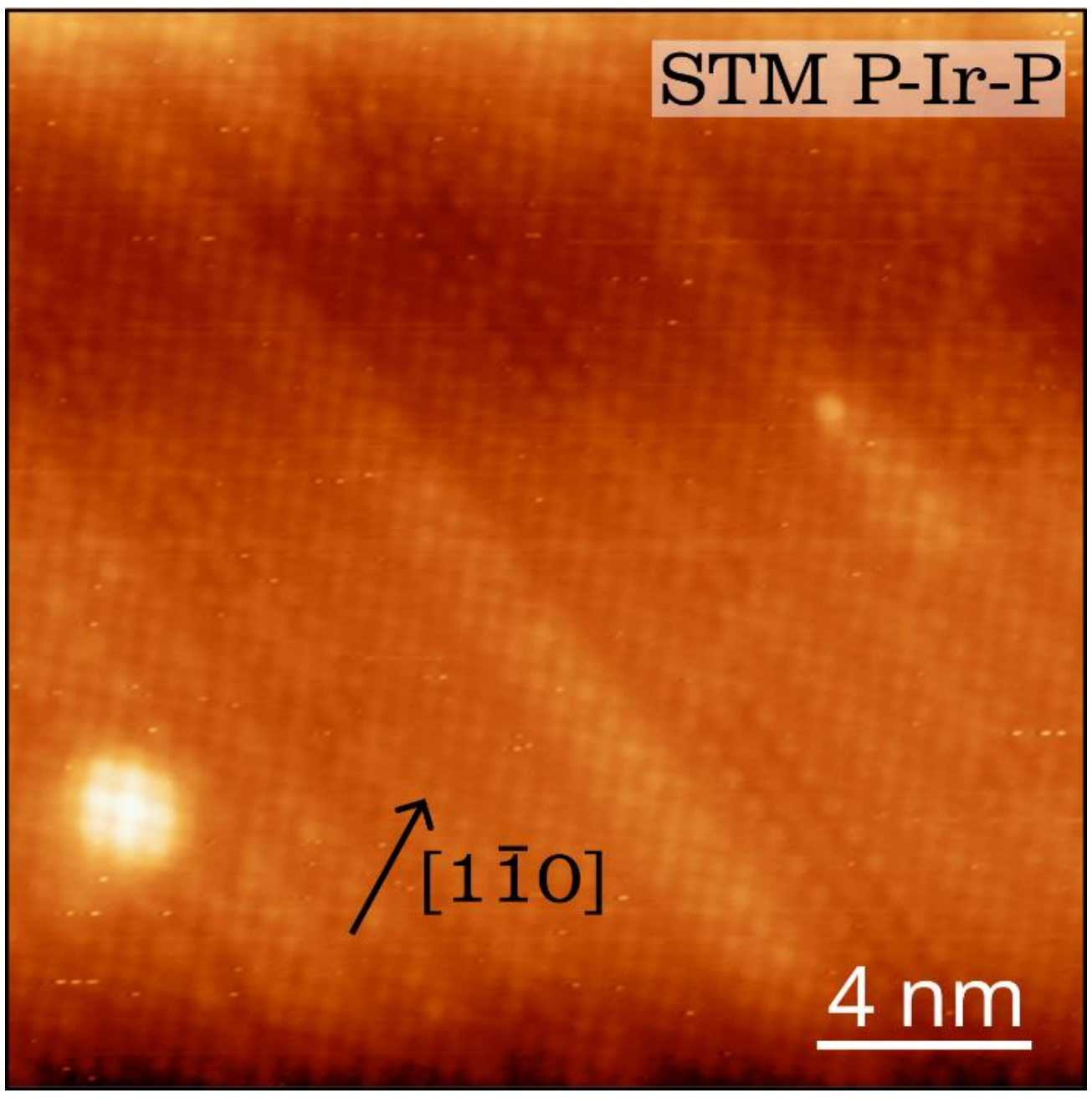

**Figure S3. Large-scale STM image of the P-Ir-P on Ir(111).** Imaging parameters: I = 200 pA, bias = 1 V, 20 × 20 $nm^2$.

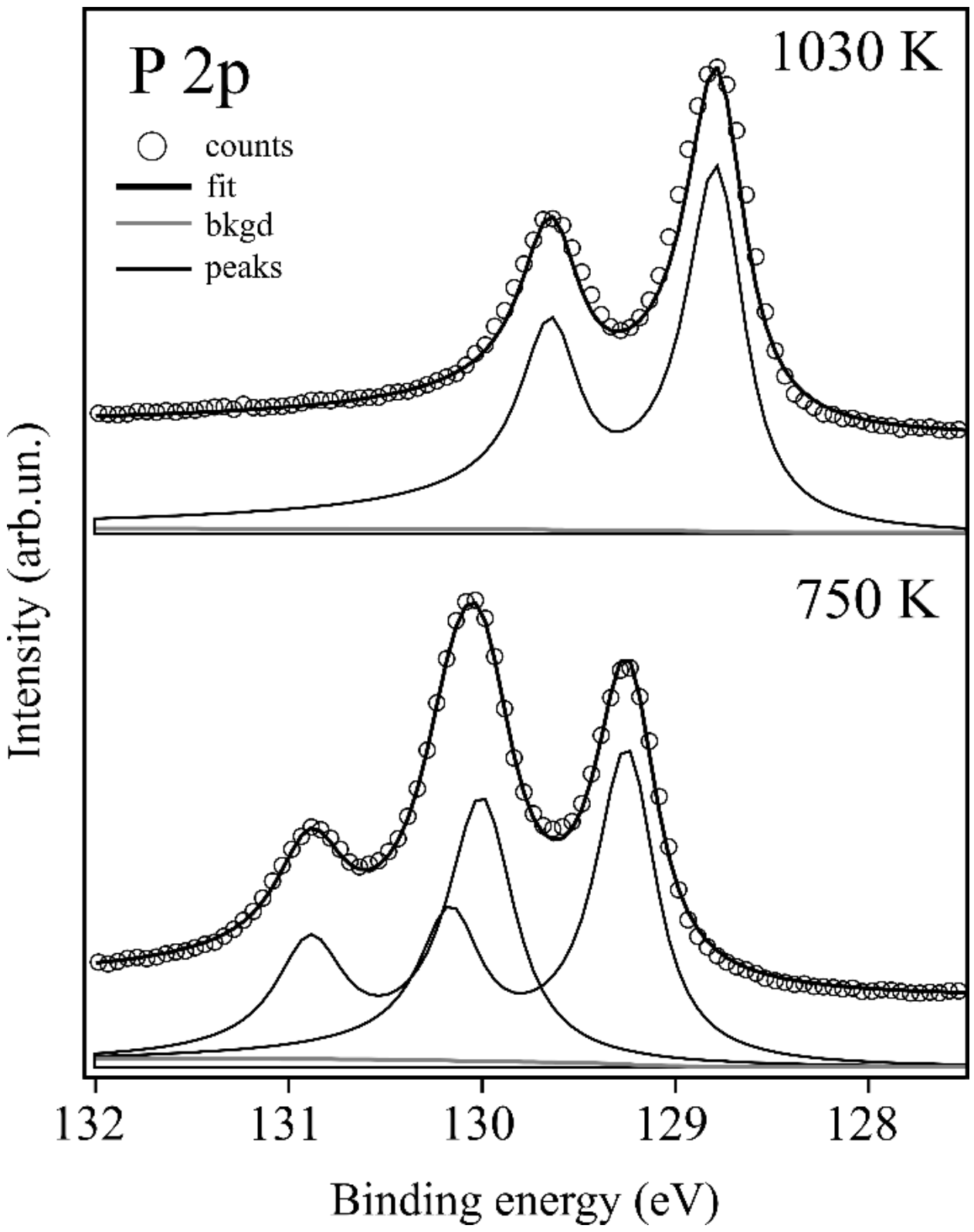

**Figure S4. P 2p XPS profiles extracted during the annealing ramp.** P 2p XPS measurements taken during the annealing ramp, at temperatures of 750 and 1030 K, indicated in Fig. 2(b) of the main text by the white dotted lines. Photon energy = 230 eV.

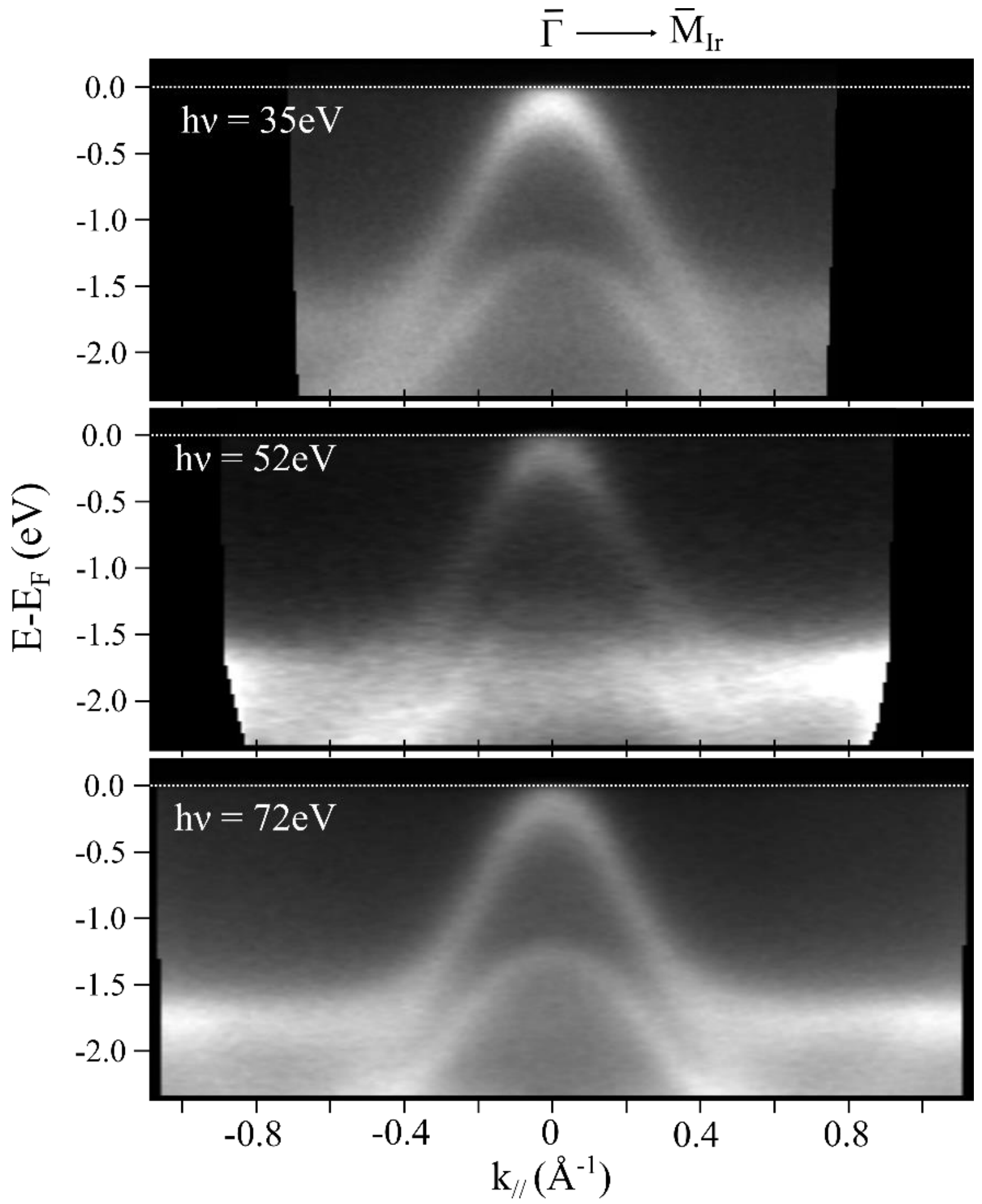

**Figure S5. Photon-energy-dependent ARPES measurements of the trilayer system.** Comparison of ARPES spectra acquired along the Γ–$M_{Ir}$ direction at photon energies of 72, 52, and 35 eV, using linearly horizontal (LH) polarization. The measurements, performed along the same direction as in Fig. 3(a), highlight the surface character of the observed electronic bands.